\documentstyle[prd,aps]{revtex}
\begin{document}
\draft
\title{SPONTANEOUS CURRENT GENERATION IN COSMIC STRINGS (DAMTP-93/34)}
\author{Patrick PETER}
\address{Department of Applied Mathematics and Theoretical Physics,\\
University of Cambridge, Silver Street, Cambridge CB3 9EW (UK),\\
{\rm and} \\
D\'epartement d'Astrophysique Relativiste et de Cosmologie\\
Observatoire de Paris-Meudon, UPR 176 CNRS, 92195 MEUDON Cedex (France)}
\date{\today}
\maketitle
\begin{abstract}
It is shown that in models including the standard electroweak theory
and for some particular values of the underlying parameters, electric
currents can be spontaneously generated in cosmic strings, without the
need of any external field (e.g., electric or magnetic) as is required
in most models. This mechanism is then shown to break spontaneously
the Lorentz invariance along the initially Goto-Nambu string. The
characteristic time needed for the current to build up is estimated
and found to lowest order to depend only on the mass of the
intermediate $W$ vector boson and the fine structure constant.
\end{abstract}

\section*{Introduction}

Cosmic strings~\cite{kibble} are linear vortex defects predicted to be
formed at a cosmological phase transition during which the vacuum
manifold would not be simply connected. The first interest in studying
them comes from the fact that since a typical grand unified theory
(GUT) predicts a few phase transitions (whose order, as far as the
strings are concerned, does actually not matter~\cite{order}), and
because the vacuum structure needed to form strings happens to be
generically realized, one can, following Vilenkin~\cite{vil-tex14},
reasonnably assume that cosmic strings have an existence probability
of at least ${1\over 2}$.  Although they are not the only possible
topological defects that could be formed in such phase transitions,
they have the advantage, as compared for instance with domain walls
and monopoles which must be somehow inflated
away~\cite{kibble,vil-rep}, to be at present compatible (as are as
well the textures~\cite{textures}) with all existent cosmological
data~\cite{hara}, while being also possibly responsible for the
formation of large scale structure~\cite{bouchet} and the
observed anisotropies in the cosmic microwave background
(CMBR)~\cite{smoot}. Most models based on these strings assume that
they are generated at the GUT phase transition, so that the
dimensionless parameter $GU$, with $G$ the Newton constant and $U$ the
energy per unit string length, giving the expected relative order of
magnitude of any gravitational effect due to these strings (e.g.,
light deflection~\cite{grav-0} or CMBR temperature
fluctuations~\cite{hara}), was assumed to be $\sim 10^{-6}$.

Another kind of strings was proposed by Witten~\cite{witten} in 1985
who pointed out the possibility that bosonic or fermionic
superconducting currents could be trapped in the strings, thereby
inducing many electromagnetic effects, such as, for instance, a new
scenario for structure formation~\cite{otw}. Shortly thereafter, it
was shown by Davis and Shellard~\cite{vorton} and independently by
Carter~\cite{meca-ring} that, although the regular strings cannot be
potentially responsible for a cosmological catastrophe (i.e., the
remnant mass density would not exceed the critical density) because of
gravitational radiation and the absence of any stabilising mechanism,
the situation was completely different for current-carrying string
loops since in the latter case, there exists centrifugally supported
equilibrium configurations (called vortons~\cite{vorton} or
rings~\cite{meca-ring}) which would overfill the
Universe~\cite{u-ring} by many orders of magnitude if they were stable
(a point which still demands further clarification and is presently
under investigation~\cite{stab-loop}), this stabilizing mechanism
being enhanced when electromagnetic corrections are taken into
account~\cite{stat-ring}, and should not be confused with the much
less efficient ``spring'', or magnetostatic support
mechanism~\cite{otw,num-witt,no-spring}.

The Witten mechanism to produce currents in cosmic strings has been
studied by many authors, interested in particular by their internal
microscopic structure~\cite{num-witt}, and who exhibited clearly the
characteristic features of what should be expected in these objects,
such as the existence of a maximum (spacelike) current (or the current
quenching phenomenon) and a phase frequency threshold (for timelike
currents). Independently, a ``macroscopic'' formalism~\cite{eos} was
derived that allows one in principle to evaluate the dynamics of any
current-carrying string configuration once its equation of state,
relating the energy per unit length $U$ to the tension $T$, is given.
This equation of state, for the Witten simple model describing
strings, and whose properties are believed to be qualitatively (if not
quantitatively) similar to more complicated (and realistic) strings
models, was indeed obtained (albeit unfortunately only numerically),
so that it has now become possible to study realistically the
cosmological importance of superconducting cosmic strings, and their
astrophysical, gravitationally induced, signature, since the structure
of the spacetime surrounding a string of any kind is also
known~\cite{grav}.

The purpose of this article is to show that there exists yet another
mechanism by which a string can become not only superconducting, but
also current-carrying without invoking any extra external field (e.g.,
electromagnetic), hence the name spontaneous current generation for
this mechanism, comparable in many respects to a similar mechanism
existing in $^3$He vortex lines~\cite{He3}. This phenomenon occurs
for particular values of the underlying microscopic parameters, when
the current-carrier field is neither a scalar nor a fermionic field
(these cases being essentially equivalent due to the two dimensional
description of the vortex), but a charged-coupled vector field such as
the intermediate $W^\pm$ in the electroweak theory. To illustrate this
mechanism in the most realistic possible way, we shall consider a
simple string-forming extension~\cite{low-mass} of the standard
electroweak theory~\cite{gsw} (this latter model being string-free
since its vacuum manifold, isomorphic to the 3--sphere $S^3$, is
simply connected). Other motivations~\cite{motivs} such as
supersymmetry or superstrings inspired models also lead, generally as
a low energy limit, to the model we wish to consider, namely that in
which an extra--$U(1)$ is gauged, this new symmetry being
spontaneously broken.  It is interesting to notice that because of the
large number of experimentally testable phenomenological consequences
of such a model, the energy scale of the symmetry breaking involved
$E_{Z'}$ say, and thus the energy per unit length of the corresponding
strings, is in fact constrained to exceed~\cite{Zprime} 300~--~500~GeV
(depending on the couplings), which is actually close to the upper
limit provided by the vorton mechanism~\cite{vorton,u-ring}, namely
$E_{Z'}\lesssim 10$~TeV.

We shall first introduce our string forming model, as well as the
vortex solutions themselves, in the simple case to begin with where no
current is flowing in the strings, i.e., the so-called
Nielsen--Olesen~\cite{NO} solutions, or Kibble type vortices. Then we
go on the spontaneous current generation itself which is shown to be
be due to an electromagnetic instability of the vacuum for massless
$W$ field at zero temperature, and finally discuss how the phenomenon
is responsible for a spontaneous breaking of the Lorentz boost
symmetry along the initially Kibble-like string.

\section{Kibble Vortices Beyond The Electroweak model.}

The electroweak theory~\cite{gsw} is based on the spontaneous breaking
of the $SU(2)_L\times U(1)_Y$ symmetry by means of an $SU(2)$ doublet
Higgs field $H$ down to the electromagnetic $U(1)$ symmetry. This
means that the vacuum has the topology of the quotient group
$SU(2)\times U(1)/U(1)$, which is isomorphic to $SU(2)$, i.e., it has
the topology of the 3--sphere and is therefore simply connected. As a
result, topologically stable cosmic strings are not present in this
model (and in fact, due to the experimental bound on the Higgs mass
$M_H \gtrsim 65$~GeV~\cite{Zprime}, even string-like solutions in
this model are dynamically unstable~\cite{Zstring,gsw-vortex}). In
order to investigate the structure of cosmic strings in a realistic
model that would take the electroweak theory into account, it is thus
necessary to modify this theory first. There are basically two
different approaches that can be followed to extend this model. The
first one consists in assuming the Higgs realisation of the symmetry
breaking not to be fundamental, and to consider instead dynamical
symmetry breaking such as in the chiral approach~\cite{chiral}
involving the $SU(2)_L\times SU(2)_R$ symmetry.  This leads to the
existence of semi-topological defects [because only one direction of
$SU(2)_R$ is actually gauged], which may be shown~\cite{Zchiral} to be
dynamically stable and moreover superconducting. The second approach,
the one we shall follow, is to regard the Higgs representation, and
thus the Higgs field $H$ itself, as fundamental, and to extend the
gauge group. It turns out that the most simple such extension one can
think of, consisting in an extra $U(1)$, also generates topologically
stable (and superconducting) cosmic strings~\cite{low-mass}.

The string-forming model we shall now examine is the following (this
section is essentially useful to fix the notation used throughout):
initially, the symmetry $SU(2)_L\times U(1)_Y\times U(1)_F$ (with $F$
the extra hypercharge) is broken down to $SU(2)_L\times U(1)_Y$ by
means of a Higgs field $\Phi$, and this is followed by the usual
electroweak phase transition. The model is minimal in the sense that
we assign a vanishing $F$ hypercharge for the $H$ field, and
symmetrically assume $\Phi$ to be an $SU(2)$ singlet. We shall
altogether neglect the fermionic sector of the model, but it may be
remarked that the hypercharge $F$, with the previous assignement made
on the Higgs fields, coincides with $B-L$ (up to a normalisation
factor absorbable in the fermionic fields) in this sector, and that
even though $U(1)_F$ is broken, the baryonic and leptonic numbers $B$
and $L$ are conserved. Also the model will be anomaly-free provided
one includes a right-handed neutrino.

We therefore start with the Lagrangian density (again, without the
fermions)
\begin{equation} {\cal L} = -{1\over 4} \vec F_{\mu\nu} \cdot \vec
F^{\mu\nu} - {1\over 4} G_{\mu\nu} G^{\mu\nu} -
{1\over 4} H_{\mu\nu} H^{\mu\nu} - (D_\mu H)^\dagger D^\mu H
- (D_\mu \Phi)^\star D^\mu \Phi - V(H,\Phi),\label{lag}
\end{equation}
where the (classical) potential between the Higgs fields is (we assume
that both phase transitions are second order so we neglect the
logarithmic corrections~\cite{order,first} in this zero temperature
effective theory, see, however, Ref.~\cite{stand-first} on that point)
\begin{equation} V(H,\Phi )= \lambda _H (H^\dagger H -
{v^2_H \over 2})^2 + \lambda _\phi (|\Phi |^2 - {v^2_\phi
\over 2})^2 + f (H^\dagger H - {v^2_H \over 2})
(|\Phi |^2 - {v^2_\phi \over 2}),\label{pot}\end{equation}
and we have set the covariant derivative
\begin{equation} D_\mu \equiv \partial _\mu - ig{\vec T\over 2}
\cdot \vec A_\mu - i g' {Y\over 2} B_\mu - i q {F\over 2} C_\mu
,\end{equation}
with $T^i$ the generators of $SU(2)_L$ in the representation of the
particle upon which the derivative acts, $g$, $g'$ and $q$ the gauge
coupling constants of $SU(2)_L$, $U(1)_Y$ and $U(1)_F$ respectively,
and the kinetic terms of the gauge vectors are expressed through
\begin{equation} F^i_{\mu\nu} = \partial_\mu A^i_\nu - \partial_\nu A^i_\mu
+ g \varepsilon^{ijk} A^j_\mu A^k_\nu,\end{equation}
\begin{equation} G_{\mu\nu} = \partial_\mu B_\nu - \partial_\nu B_\mu \
\ \ , \ \ \ H_{\mu\nu} = \partial_\mu C_\nu - \partial_\nu C_\mu .
\end{equation}
The Higgs doublet is understood as $H = (H^+,H^0)$, and
its vacuum expectation value (VEV) is experimentally known to be
$\sqrt{\langle |H|^2 \rangle _0 } = v_H /\sqrt{2}\simeq174$~GeV.

We are now interested in a vortex solution of this model, of the kind
proposed by Nielsen and Olesen~\cite{NO}, for the $\Phi$ field.
Since we are concerned by classical solutions, it is necessary that we
fix the gauges. For the string-forming fields, there is a particularly
convenient gauge choice: if the vortex solution is taken to be aligned
along the $z$ axis (which is always possible since the curvature
effects can be locally neglected), we can choose a cylindrical
coordinate system, and in this system, the phase of the $\Phi$ field
is identified with the angular coordinate $\theta$. The
Nielsen--Olesen vortex solution then takes the simple form
\begin{equation} \Phi = \varphi (r) \exp{i n \theta} ,\label{no}
\end{equation}
with $n$ the winding number~\cite{kibble,order,vil-rep}.

Let us now turn to the electroweak fields. Because of the disjoint
structure of the initial invariance, we have not lost any freedom in
going to the vortex gauge. We can thus choose the most convenient
gauge with regard to the subsequent interpretation, namely the unitary
gauge, in which only the neutral component of $H$ is considered:
$H = [0, h(r)/\sqrt{2}]$. Before going any further in the resolution
of the Euler--Lagrange equations for this system, we wish to examine
in more details what occurs in the strings core.

The string solution is defined as the set of points in space where
$\Phi =0$. Moreover, the vacuum (or the false vacuum in the case of
the strings core) should represent a minimum of the
potential~(\ref{pot}). Varying this potential for $h$ and $\varphi$,
we see that the extremization yields two differents possibilities,
namely, far from the strings core, i.e., in the usual vacuum,
\begin{equation} h=v_h \ \ \hbox{and} \ \
\varphi = v_\phi /\sqrt{2},\label{reg-vac}\end{equation}
whereas in the strings core with $\varphi =0$, then $h$ should
satisfy (not taking the kinetic terms into account for the moment)
\begin{equation} h^2 = v_H^2 + f{v_\phi ^2\over
\lambda _H},\label{S-vac}\end{equation}
from which we can conclude that two cases may occur in principle. The
first case, already studied elsewhere~\cite{low-mass}, is for
$f>f_{crit}$, with
\begin{equation} f_{crit} = -\lambda _H \left( {v_H \over
v_\phi }\right)^2 ,\label{f:crit}\end{equation}

\noindent which corresponds to a shift in the $SU(2)$ doublet
Higgs VEV at $r=0$. The second case, to which we now turn definitely,
is for $f\leq f_{crit}$. If the underlying parameters are such that
this inequality is satisfied, then there is no real solution to
Eq.~(\ref{S-vac}). Thus, one finds that the real minimum of the
potential is now at $h=0$ as long as $\varphi
\leq \varphi _{min}$, where $\varphi ^2 _{min} = v^2 _\phi /2 -
\lambda _H v^2_H /|f|$ [given by inserting a nonzero value for
$\sigma$ into Eq.~(\ref{S-vac}) and solving for $h=0$].

Fig.~1 illustrates the internal string structure which is obtained
when the kinetic terms are included. This figure represents a solution
of the field equations derived from the Lagrangian~(\ref{lag}) under
the gauge assumptions and with the vortex solution~(\ref{no}), with
zero vector fields $A_{i\mu}$ and $B_\mu$.This solution was obtained
by means of a successive over relaxation method~\cite{adler}, and the
distances are in units of the inverse $\Phi$ mass $(\lambda _\phi
v_\phi )^{-1}$. More details concerning the numerical procedure
itself and the stability of the solution can be found in
Ref.~\cite{low-mass}, but here, and in particular in the next section,
we shall be mainly interested in what occurs close to the strings
core, namely the symmetry restoration. For the time being, let us just
remark that since the Higgs field $h$ is real, there is no
associated current with it, so the fact that it be trapped in the
string, its VEV varying from $r=0$ to $r\to\infty$, merely changes the
actual value of the string's energy per unit length, but otherwise
does not break the Lorentz boost invariance along the string.
Therefore, setting the stress-energy tensor in the form
\begin{equation} T^{\mu\nu} = U u^\mu u^\nu - T v^\mu v^\nu
,\label{Tmn}\end{equation}

\noindent with $u$ and $v$ two unit timelike and
spacelike vector respectively, tangent to the strings worldsheet, $U$
being the energy per unit length and $T$ the tension, the Lorentz
invariance requires, whether there is a Higgs condensate or not, that
the equation of state be that of Goto--Nambu, i.e.,
$U=T=$Cte~\cite{vil-rep,eos}. This is important because once the
current-generation mechanism which we will investigate in the next
section has been at work, this degeneracy in the stress-energy tensor
eigenvalues is spontaneously raised, so the Lorentz invariance is
spoiled.

\section{Intermediate Boson Instability.}

The electroweak vacuum surrounding a cosmic string of the kind we just
investigated is in fact not stable. This can be seen as follow: in the
standard vacuum, the $W^\pm$ particles are charged and massive because
of the Higgs field $H$ VEV. Now, close to the string core, as we have
just seen, this VEV actually vanishes so the $W^\pm$ particles become
charged and massless. As a result, they can be created by pair through
any fluctuation of the electromagnetic field, but since they are
charged, they can actually be considered themselves as the sources for
these electromagnetic fluctuations. More precisely, as will be shown
in this section, fluctuations in the $W$ field yield a corresponding
nonvanishing $A$ and $Z$, with nonzero gradients. This implies nonzero
electric and magnetic fields which are used as negative masses for the
$W$ particles. The vacuum surrounding a cosmic string is thus unstable
and there is a spontaneous current generation in the form of $W$
flowing along the strings.

To see how this phenomenon actually occurs, let us concentrate on the
stress energy tensor $T^{\mu\nu}$ given by
\begin{equation} T^{\mu\nu}=-2g^{\mu\alpha}g^{\nu\sigma}{\delta {\cal
L} \over \delta g^{\alpha\sigma}} +g^{\mu\nu}{\cal L},\end{equation}

\noindent and in particular the energy density ${\cal U}=T^{tt}$.
Setting as usual~\cite{num-witt} $Q(r)=n-{1\over 2}q C_\theta$,
$W^\pm_\mu = (A_{1\mu} \mp i A_{2\mu})/\sqrt{2}$, $Z_\mu =c
A_{3\mu}-sB_\mu$, $A_\mu = s A_{3\mu} + cB_\mu$, $s\equiv \sin
\theta_W$, $c\equiv \cos \theta_W$, $\tan \theta _W \equiv g'/g$, and
using the vortex ansatz~(\ref{no}) gives, if one considers only a
configuration where radial electric and orthoradial magnetic fields
are present (i.e., with $A_z (r)$, $A_t (r)$, $Z_z (r)$ and $Z_t (r)$
the only nonvanishing components of the photonic and the $Z$ fields)
\begin{eqnarray} {\cal U} &=& {1\over 2} h'^2 + \varphi '^2 +
{\varphi^2 Q^2 \over r^2} +{2 Q'^2 \over r^2 q^2} + {\lambda_H\over 4}
(h^2 - v_H^2 )^2 +\lambda _\phi (\varphi ^2-{v_\phi^2\over 2})^2 +
{1\over 2}f (h^2 - v_H^2 ) (\varphi ^2-{v_\phi^2\over 2}) \nonumber\\
&+&2 W^+_{\mu t}W^{-\mu}_{\ \ \ t} + {1\over 2} W^+_{\mu\nu} W^{-\mu\nu} +
{1\over 2} (A_z'^2 + A_t'^2 + Z_z'^2 + Z_t'^2 ) + {g^2\over 8c^2}h^2
(Z_z^2 + Z_t^2) \nonumber\\
&+&2ig\left\{ W^+_{\mu t}\left[ W^- _t
(sA^\mu + c Z^\mu) - W^{-\mu} (sA_t + c Z_t) \right] + W^-_{\mu t}
\left[ W^{+\mu} (sA_t + c Z_t) - W^+_t (sA^\mu + c Z^\mu)
\right]\right\} \nonumber \\
&+&{1\over 2}ig\left\{ W^+_{\mu\nu}\left[ W^{-\nu}
(sA^\mu + c Z^\mu) - W^{-\mu} (sA^\nu + c Z^\nu) \right]
+ W^-_{\mu\nu} \left[ W^{+\mu} (sA^\nu + c Z^\nu) - W^{+\mu} (sA^\nu +
c Z^\nu) \right] \right\} \nonumber\\
&+&{1\over 4}g^2h^2(|W_r|^2 + |W_z|^2 +
{1\over r^2} |W_\theta |^2 + |W_t|^2 ) \nonumber\\
&+&ig[(sA'_z + cZ'_z)(W_r^- W^+_z - W_z^-W^+_r)
+(sA'_t + cZ'_t)(W_r^- W^+_t - W_t^-W^+_r)] \nonumber\\
&-& g^2 \Big\{ \Big| (sA_z + cZ_z) W^+_z + (sA_t+cZ_t)W^+_t \Big| ^2
\nonumber\\
&+& {1\over 2}(W^+_r W^-_t - W^-_r W^+_t )^2
+{1\over 2}(W^+_z W^-_t - W^-_z W^+_t )^2 +
{1\over 2}(W^+_r W^-_z - W^-_r W^+_z )^2 \nonumber\\
&+& {1\over 2 r^2} \left[
(W^+_r W^-_\theta - W^-_r W^+_\theta )^2
+ (W^+_z W^-_\theta - W^-_z W^+_\theta )^2 + (W^+_t W^-_\theta -
W^-_t W^+_\theta )^2 \right] \nonumber\\
&-& (|W_r|^2 + |W_z|^2 + {1\over r^2} |W_\theta |^2 + |W_t|^2 )
\left[ (sA_z + cZ_z)^2 + (sA_t+cZ_t)^2 \right] \Big\} ,
\label{bigU}\end{eqnarray}

\noindent where a prime means differentiation with respect to the
radial coordinate $r$. Let us now consider the quadratic terms in
$W^\pm$ that are present in Eq.~(\ref{bigU}), i.e., the effective mass
matrix $M_{ij}^2$, defined by
\begin{equation} M_{ij}^2 \equiv 2 {\delta {\cal U} \over \delta W_i^+
\delta W_j^- } \Big| _{W_i^\mu = 0 = A^\mu = Z^\mu}.
\label{Mmatrix}\end{equation}

\noindent Because of the coupling between the $W$ fields and the
gradients of the photon and the $Z$ fields, this mass matrix is in
fact nondiagonal, and one can easily derive the
eigenvalues as
\begin{equation} m_1 ^2 = {1\over 2} g^2 h^2 ,\label{m1}\end{equation}
\begin{equation} m_2 ^2 = {1\over 2} g^2 h^2 + 2
g\sqrt{(sA'_z+cZ'_z)^2 + (sA'_t+cZ'+t)^2},
\label{m2}\end{equation}
\begin{equation} m_3 ^2 = {1\over 2} g^2 h^2 -
2 g\sqrt{(sA'_z+cZ'_z)^2 + (sA'_t+cZ'+t)^2},
\label{m3}\end{equation}

\noindent where, to first order in $g$, the eigenvectors are
\begin{equation} W^-_1 = {1\over \sqrt{\beta_z^2 + \beta_t ^2}}
(\beta_t W^-_z - \beta_z W^-_t) ,\end{equation}
\begin{equation} W^-_2 = {1\over \sqrt{2}}\left[ W^-_r - {i\over
\sqrt{\beta_z^2 + \beta_t ^2}} (\beta_z W^-_z + \beta_t W^-_t)
\right],\end{equation}
\begin{equation} W^-_3 = {1\over \sqrt{2}}\left[ W^-_r + {i\over
\sqrt{\beta_z^2 + \beta_t ^2}} (\beta_z W^-_z + \beta_t W^-_t)
\right],\label{W3}\end{equation}

\noindent and we have set $\beta_a = g(sA'_a + c Z'_a)$, $a=z,t$ the
radial (respectively orthoradial) component of the electric (resp.
magnetic) field.

It can be remarked on Eqs.~(\ref{m1}), (\ref{m2}) and~(\ref{m3}) that
$W_3$ has a nonpositive definite mass, which is not the case for all
other components of this gauge field. Therefore, since we are seeking
a minimum energy configuration, it seems safe to assume $W_1 = W_2 = 0
= W_\theta$. Setting, for simplicity, $W^+_3 \equiv W$, one has
\begin{equation} W^+_r = {1\over \sqrt{2}} W,
\label{Wr}\end{equation}
\begin{equation} W^+_z = {i\beta_z\over \sqrt{2(\beta_z^2 +
\beta_t ^2)}}W ,\label{Wz}\end{equation}
\begin{equation} W^+_t = {i\beta_t\over \sqrt{2(\beta_z^2 +
\beta_t ^2)}}W ,\label{Wt}\end{equation}

\noindent where now $\beta_z$ and $\beta_t$ are arbitrary. We recover
the previous results~\cite{low-mass} $W^-_r =
\pm i W^-_z$ or $W^-_r = \pm i W^-_t$ for the purely magnetic or
electric cases respectively (i.e., when only one component of $A$ and
$Z$ is explicitely considered).

Under the assumptions given by Eqs.~(\ref{Wr}), (\ref{Wz})
and~(\ref{Wt}), it is now simple to derive the effective potential for
the $W$ field, namely:
\begin{equation} V(W) = {1\over 2} m_3 ^2|W |^2 +
g^2 |W |^4 ,\label{potU}\end{equation}

\noindent so this field effectively behaves
as a Higgs ``scalar'' in the region of parameter space where its
squared mass is negative. This is indeed possible close to the strings
core since there, as we have seen, still in the case where
$f<f_{crit}$, the Higgs field $h$ vanishes, so to first order in $g$,
Eq.~(\ref{m3}) implies that $m_3^2$ is actually negative. Thus,
any fluctuation in $W$, the latter being coupled with the
photon, will generate a small fluctuation in $A_z$, $A_t$, $Z_z$ and
$Z_t$, which, if the initial perturbation was axisymmetric, will
produce an electric or a magnetic field. Since $W$ is
effectively massless in the core of the string, the energy in the
electromagnetic perturbation is already sufficient to create a pair
$W^+ W^-$, which can in turn be seen as the source for
the electromagnetic fields. Since these electromagnetic fields are
necessary to support the $W$ condensate in the strings core,
one is led to the conclusion that a current has been spontaneously
generated. We shall now investigate in more details this mechanism.

\section{Spontaneous Current Generation.}

To exhibit the instability of the electroweak vacuum surrounding a
Nielsen-Olesen string~(\ref{no}), we turn to the Euler-Lagrange
equations derivable from the Lagrangian~(\ref{lag}), which we expand
to first order in the coupling constant $g$ and to lowest order in the
various fields involved to consider the case of a perturbation in
$W_3$ [given by Eq.~(\ref{W3})]. For the photon, we have
\begin{equation} \partial_\beta \left\{ r\left[ A^{\alpha\beta} + igs
\left( W^{-\alpha} W^{+\beta} - W^{-\beta} W^{+\alpha} \right)\right]
\right\} = rigs \left( W^{-\alpha\beta} W^+_\beta - W^{+\alpha\beta}
W^-_\beta \right),\label{Apert}\end{equation}

\noindent a similar equation applying for the $Z$ field with $\sin
\theta _W$ replaced by $\cos \theta _W$, so also similar conclusions
can actually be drawn for both fields, and
\begin{equation} \partial_\beta \left\{ r\left[ W^{-\alpha\beta} - ig
W^{-\alpha} (sA^\beta+cZ^\beta) +ig W^{-\beta} (sA^\alpha +cZ^\alpha)\right]
\right\} = rigs \left[ W^{-\beta\alpha} (sA_\beta +cZ_\beta)
+(sA^{\beta\alpha} +cZ^{\beta\alpha})W^-_\beta \right],
\label{Wpert}\end{equation}

\noindent close to the strings core, in the symmetry restoration
region where $h=0$. In fact, because there, $h=0$, neglecting a
possible backreaction due to outer region couplings, we see that the
background Nielsen-Olesen string is essentially unaffected by the
inclusion of the $W$, $A$ and $Z$ fields.

We shall now examine a perturbation in $W_3 = W$ in the form
$W = |W (r)| e^{i \omega t}$. Inserting this form into
Eq.~(\ref{Apert}) yields
\begin{equation} {\partial^2 A_z \over \partial z \partial r}
-{\partial ^2 A_t \over \partial t \partial r} - {\partial ^2 A_r
\over \partial z^2} + {\partial ^2 A_r \over \partial t^2} = 0,
\label{Ar} \end{equation}
\begin{equation} {\partial^2 A_r \over \partial z \partial r} -
{\partial^2  A_z \over \partial r^2} -{\partial^2 A_t \over \partial z
\partial t} + {\partial^2 A_z \over \partial t^2} +{1\over r} {\partial
A_r \over \partial z} - {1\over r}{\partial A_z\over \partial r} =
{gs\beta_z \over \sqrt{\beta^2_z +\beta^2_t}} \left[ \left( {\beta _t
\omega \over \sqrt{\beta^2_z +\beta^2_t}} - {1\over r} \right)
|W |^2 - 3 |W | {d|W |\over dr} \right] ,
\label{Az1}\end{equation}
\begin{equation} {\partial^2 A_r \over \partial r \partial t} -
{\partial^2  A_t \over \partial r^2} +{\partial^2 A_z \over \partial z
\partial t} - {\partial^2 A_t \over \partial z^2} +{1\over r} {\partial
A_r \over \partial t} - {1\over r}{\partial A_t\over \partial r} =
{gs\beta_t \over \sqrt{\beta^2_z +\beta^2_t}} \left[ \left( {\beta _t
\omega \over \sqrt{\beta^2_z +\beta^2_t}} - {1\over r} \right)
|W |^2 - 3 |W | {d|W |\over dr} \right] ,
\label{At1}\end{equation}

\noindent equations in which we shall assume for now on that $\partial _z$
is ignorable, and we will denote the differentiation with respect to
the radial coordinate $r$ by a prime, whereas a dot will mean a
derivative with respect to time. We shall also work in the Lorentz
gauge $\nabla _\mu A^\mu = 0$ since in this gauge the equations for
the various components of $A$ are decoupled. Actually, the gauge
condition together with the field equation~(\ref{Ar}) with $\partial
_z$ ignorable yields, upon differentiation with respect to $r$
\begin{equation} \dot A'_r + {1\over r} \dot A_r = \ddot A_t
,\label{gauge}\end{equation}

\noindent and inserting this into Eq.~(\ref{Ar}) yields
\begin{equation} \ddot A_r = A_r '' + {1\over r} A'_r - {1\over r^2}
A_r ,\end{equation}

\noindent so that setting $A_r = f(t) g(r)$ gives
\begin{equation} {\ddot f\over f} = {\Delta_2 g\over g} - {1\over r^2}
= - \alpha ^2 ,\end{equation}

\noindent with $\alpha$ a constant and $\Delta_2$ the two dimensionnal
Laplacian in the transverse plane. So $f\propto e^{i\alpha t}$, and
the radial dependent part obeys a Schr\"odinger equation
\begin{equation} -\Delta_2 g + {1\over r^2} g = \alpha ^2 g
,\label{schro}\end{equation}

\noindent with a positive definite potential $r^{-2}$. Therefore, the
eigenvalues $\alpha ^2$ of Eq.~(\ref{schro}) are all positive and the
string state is stable against $A_r$ perturbations. Thus, neglecting
$A_r$ in the resulting stationnary configurations is justified and we
shall consistently set $A_r = 0$ in what follows.

Inserting Eq.~(\ref{gauge}) into Eq.~(\ref{At1}) is the last step
toward the definite equations describing the dynamics of $A_z$ and
$A_t$ when a perturbation in $W_3$ is applied, and we find
\begin{equation} \ddot A_z - A_z '' - {1\over r} A_z ' =
{gs\beta_z \over \sqrt{\beta^2_z +\beta^2_t}} \left[ \left( {\beta _t
\omega \over \sqrt{\beta^2_z +\beta^2_t}} - {1\over r} \right)
|W |^2 - 3 |W | {d|W |\over dr} \right] ,
\label{Az}\end{equation}
\begin{equation} \ddot A_t - A_t '' - {1\over r} A_t ' =
{gs\beta_t \over \sqrt{\beta^2_z +\beta^2_t}} \left[ \left( {\beta _t
\omega \over \sqrt{\beta^2_z +\beta^2_t}} - {1\over r} \right)
|W |^2 - 3 |W | {d|W |\over dr} \right] ,
\label{At}\end{equation}

\noindent out of which the spontaneous current mechanism can be
clearly exhibited.

The linearized field equations~(\ref{Az}) and~(\ref{At}) have a first
immediate consequence, namely that $\beta_t A_z = \beta _z A_t$. It
turns out that this relation is still valid when the whole set of
classical field equations are used, so the overall field configuration
is in fact determined by the value of the ratio $\beta_z /
\beta_t$. We will return to that point later. Moreover,
Eqs.~(\ref{Az}) and~(\ref{At}), being inhomogeneous because of the
source term due to the $W$ field [a direct consequence of the
nonabelian nature of $SU(2)\times U(1)$], the configuration $(A_z = 0$
and $A_t = 0)$ is not solution of the field equations, and more
generally there is also no solution with vanishing gradient. But this
is precisely the condition for the squared mass $m_3^2$ to be
nonpositive definite. Thus, we know that there exist unstable modes in
Eq.~(\ref{Wpert}). These modes will grow exponentially, as we shall
now show, as will also $A_t$ and $A_z$, until they reach an
equilibrium configuration where the quadratic terms become large
enough to stop the instability.

To examine the $W$ instability, we note first that Eq.~(\ref{Wpert})
is not well adapted since it was written for the usual components of
$W^\mu$. Instead, we write an equivalent linearized action [obtained
by retaining only the lowest order tems in the Lagrangian~(\ref{lag})]
for the field $W=W_3^+$ alone, namely
\begin{eqnarray} S = \pi\int r\,dr\,dt\,&\Big\{& {\beta_t^2 -
\beta_z^2 \over \beta_z^2 + \beta_t^2} |W'|^2 - {2\beta_z^2 +
\beta_t^2 \over  \beta_z^2 + \beta_t^2} |\dot W|^2 +{i\beta_t\over
\sqrt{\beta_z^2 + \beta_t^2}} (W'\dot W^\star - \dot W W'^\star )
\nonumber\\
&-&g(sA_z+cZ_z)\big[ {\beta_z \over \sqrt{\beta_z^2 + \beta_t^2}}
(W'^\star W + W^\star W') + {i\beta_z\beta_t\over\beta_z^2 +
\beta_t^2} (\dot W W^\star - \dot W^\star W) \big] \nonumber\\
&+&g(sA_t+cZ_t)\big[ {\beta_t \over \sqrt{\beta_z^2 + \beta_t^2}}
(W'^\star W + W^\star W') + i{2\beta_z^2 + \beta_t^2\over\beta_z^2 +
\beta_t^2} (\dot W W^\star - \dot W^\star W) \big] \nonumber\\
&+&{2g\over\sqrt{\beta_z^2 +\beta_t^2}} \big[ \beta_z (sA'_z+cZ'_z) -
\beta_t (sA'_t+cZ'_t) \big] |W|^2 \Big\} \end{eqnarray}

\noindent whose variations yield the following Euler-Lagrange equation:
\begin{equation} a(W''+{1\over r}W') +(\delta'+{1\over r}\delta) W -
i {c\over r} \dot W + b \ddot W - i\dot \epsilon W - 2 i\epsilon \dot W
= \sigma W ,\label{Wfin}\end{equation}

\noindent where we have set $a=(\beta_t^2-\beta_z^2)/(\beta_z^2 +
\beta_t^2)$, $b=-(2\beta_z^2 + \beta_t^2)/(\beta_z^2 +\beta_t^2)$,
$c=\beta_t /\sqrt{\beta_z^2 +\beta_t^2}$
\begin{equation} \delta (r,t) = -{g\over \sqrt{\beta_z^2 +\beta_t^2}}
\big[ \beta_z (sA_z+cZ_z) - \beta_t (sA_t + cZ_t)\big],\end{equation}
\begin{equation} \epsilon (r,t) = {g\beta_z \beta_t \over \beta_z^2 +
\beta_t^2} (sA_z+cZ_z) - g{2\beta_z + \beta _t \over \beta_z^2 +
\beta_t^2} (sA_t+Z_t),\end{equation}

\noindent and
\begin{equation} \sigma (r) = {2 g \over \sqrt{\beta_z^2 +
\beta_t^2}} \big[ \beta_z (sA'_z+cZ'_z) + \beta_t (sA'_t+cZ'_t) \big]
.\end{equation}

Although it is impossible to solve Eq.~(\ref{Wfin}), some information
regarding the transition from the nonconducting state to the
current-carrying state can be obtained from it if one makes an
``adiabatic'' hypothesis, namely assumes that the transition is slow
enough that the time derivative in the gauge fields can be neglected
compared to their spatial gradients. In this hypothesis, setting
$W=\xi (t) \rho (r)$ into Eq.~(\ref{Wfin}) yields
\begin{equation} a {\rho '' + \rho' /r \over \rho} + \delta' +{1\over
r} \delta - \sigma = i {\dot \xi \over \xi} ({c\over r} + 2\epsilon )
- b {\ddot \xi \over\xi} \equiv - \zeta (r) ,\end{equation}

\noindent where the function $\zeta$ depends on the radial
coordinate $r$ only. Thus, the oscillatory modes $\xi \propto e^{i\omega
t}$ satisfy a dispersion relation
\begin{equation} \omega ^2 - {1\over b} ({c\over r} + 2\epsilon )
\omega + {\zeta \over b} = 0 ,\label{modes}\end{equation}

\noindent whose solutions have an imaginary part for $\zeta > {1\over
4b} (c/r + 2\epsilon)^2$, i.e., in the low frequency limit (thereby
justifying the ``adiabatic'' hypothesis). In the limit of zero
frequency ($\omega \to 0$), we can estimate roughly, in order of
magnitude, the expected value of the timescale necessary for the
string to become current-carrying (namely $\tau \sim \omega^{-1}$):
assuming the current-carrier field to have an
amplitude~\cite{num-witt} $|W| \sim M_W$, and taking $r_0$ to be the
typical distance over which the fields vary, then Eqs.~(\ref{Az})
and~(\ref{At}) give $A_z \sim gsr_0 M_W^2$, and we find
\begin{equation} \tau \sim (eM_W)^{-1} , \label{time}\end{equation}

\noindent i.e., a time independent of the string thickness, an
expected result since the background string and the electroweak fields
are decoupled in the core (again, neglecting backreaction). It should
be mentionned that for coupling values $f>f_{crit}$, the same
mechanism actually applies, but that in this case, the initial
configuration and only metastable and although the current will
definitely be spontaneously generated, it will be through tunnelling.
As a result, the life time of the noncurrent-carrying configuration
is in fact increased by a factor depending on the strings
thickness and the value of $h$ at $r=0$, these giving the order of
magnitude of the expected potential barrier and its width: the
standard WKB approximation then gives an extra exponential factor in
Eq.~(\ref{time}).

\section{Internal String Structure.}

The spontaneous current generation mechanism we have just discussed
has in fact many interesting consequences, including, we believe,
cosmological (notably in the framework of the vorton
problem~\cite{vorton} which becomes even more unavoidable in this
context), and in this section, we wish to emphasize a particular
effect, namely that generating a current this way breaks the Lorentz
boost invariance along the string spontaneously. The basic
reason that this occurs is that the particle that
gets trapped in the string is in fact a $W^\pm _\mu$, i.e., a
vectorial particle, and a nonvanishing VEV for a vector picks a
privileged direction in space time. Also we shall exhibit the internal
microscopic structure of the string and compare it with what is
obtained in the simple Witten~\cite{witten,num-witt} bosonic toy
model.

As was already said, the current generation phenomenon will
spontaneously break the Lorentz boost symmetry along the string:
before the $W$ condenses in the strings core, the energy per unit
length and the tension are both equal so a boost in the $z$ direction
does not change the physics of the system. Now, when a perturbation in
$W$ is applied, as we have just seen, the $W$ and $A$ fields VEV
increase exponentially, so the degeneracy of the stress energy tensor
is raised exponentially as times passes by, until again the system
reaches a stationnary configuration. It is therefore an actual
spontaneous mechanism because it does not require any external field
and it needs not even exist in the expanding Universe (as is the
case for the usual symmetry breaking Higgs mechanism).

A stationnary configuration obtained this way consists in a $W$ field
together with, as argued before, any value for the ratio $\beta_z /
\beta_t$. In fact, it turns out that the only thing to know to
determine (nearly) entirely the configuration is whether this ratio is
less or greater than unity, for once this is known, it suffices to
apply a boost along the string to remove one of the field $A_z$ or
$A_t$. As a result, the only interesting cases are the magnetic case
for which one can always set $\beta_t = 0$ and $A_t=Z_t=0$, the
electric case having $\beta_z = 0$ and $A_z=Z_z=0$, and the null or
lightlike case with $\beta_z = \beta_t \equiv \beta$, $A_z=A_t\equiv
A$ and $Z_z=Z_t\equiv Z$, as can be seen on Eqs.~(\ref{Az}) and
(\ref{At}). It can then be shown~\cite{num-witt,low-mass} that the
most general configuration will be
\begin{equation} W = \Upsilon (r) e^{i \psi (z,t)} ,\end{equation}

\noindent where, without lack of generality~\cite{num-witt},
the phase function $\psi$ could be chosen as $\psi = \omega t - k z$.
The ``energy per unit length'' is then
\begin{eqnarray} \tilde U = 2\pi \int r\,dr\,&\Big\{& {1\over 2}h'^2
+\varphi '^2 +{\varphi ^2 Q^2 \over r^2} + {2 Q'^2 \over r^2 q^2} +
{\lambda_H \over 4} (h^2 - v_h^2)^2 + \lambda_\phi (\varphi ^2 -
{v_\phi ^2 \over 2})^2 + {1\over 2} f (h^2 - v_h^2) (\varphi ^2 -
{v_\phi ^2 \over 2}) +{1\over 2}\Upsilon'^2\nonumber \\
&+& {1\over 2} [(\partial _z \psi )^2 + (\partial _t\psi )^2 + {(\beta
_t \partial_z\psi + \beta_z\partial_t\psi )^2\over \beta^2_z
+\beta^2_t } ] \Upsilon ^2 + {\beta_z\partial_z\psi +\beta_t\partial_t
\psi \over \sqrt{\beta^2_z+\beta^2_t}} \Upsilon \Upsilon ' +{1\over 2}
(A'^2_z + A'^2_t + Z'^2_z + Z'^2_t) \nonumber\\
&+&{g^2\over 8c^2} h^2 (Z_z^2
+Z_t^2) +{1\over 4} g^2 h^2 \Upsilon ^2 +{1\over 2} g^2 \Upsilon ^4 + g^2
\Upsilon ^2 [(sA_z+cZ_z)^2 + (sA_t+cZ_t)^2]\nonumber\\
&+& g(sA_z+cZ_z)\big\{ {\beta_z\Upsilon\Upsilon'\over
\sqrt{\beta^2_z+\beta^2_t}} + \big[ \partial_z\psi + {\beta_t \over
\beta^2_z+\beta^2_t} (\beta_t \partial_z\psi + \beta_z\partial_t\psi)
\big] \Upsilon^2 \big\} \nonumber\\
&+& g(sA_t+cZ_t)\big\{ {\beta_t\Upsilon\Upsilon'\over
\sqrt{\beta^2_z+\beta^2_t}} + \big[ \partial_t\psi - {\beta_z \over
\beta^2_z+\beta^2_t} (\beta_t \partial_z\psi + \beta_z\partial_t\psi)
\big] \Upsilon^2 \big\} \nonumber\\
&-&{g^2\Upsilon^2\over 2(\beta^2_z+\beta^2_t)} \big[ (sA_z+cZ_z)\beta_z +
(sA_t+cZ_t) \beta_t \big]^2 - {g\Upsilon^2\over
\sqrt{\beta^2_z+\beta^2_t}} \big[ (sA'_z+cZ'_z)\beta_z +
(sA'_t+cZ'_t) \beta_t \big] \Big\} ,\\
\label{Utilde}\end{eqnarray}

\noindent and the ``tension''
\begin{eqnarray} \tilde T = 2\pi \int r\,dr\,&\Big\{& {1\over 2}h'^2
+\varphi '^2 +{\varphi ^2 Q^2 \over r^2} + {2 Q'^2 \over r^2 q^2} +
{\lambda_H \over 4} (h^2 - v_h^2)^2 + \lambda_\phi (\varphi ^2 -
{v_\phi ^2 \over 2})^2 + {1\over 2} f (h^2 - v_h^2) (\varphi ^2 -
{v_\phi ^2 \over 2}) -{1\over 2}\Upsilon'^2\nonumber \\
&-& {1\over 2} [(\partial _z \psi )^2 + (\partial _t\psi )^2 + {(\beta
_t \partial_z\psi + \beta_z\partial_t\psi )^2\over \beta^2_z
+\beta^2_t } ] \Upsilon ^2 - {\beta_z\partial_z\psi +\beta_t\partial_t
\psi \over \sqrt{\beta^2_z+\beta^2_t}} \Upsilon \Upsilon ' -{1\over 2}
(A'^2_z + A'^2_t + Z'^2_z + Z'^2_t)\nonumber\\
&-&{g^2\over 8c^2} h^2 (Z_z^2 +Z_t^2) - {1\over 2} g^2 \Upsilon ^4
- g(sA_z+cZ_z)\big\{ {\beta_z\Upsilon\Upsilon'\over
\sqrt{\beta^2_z+\beta^2_t}} + \big[ \partial_z\psi - {\beta_t \over
\beta^2_z+\beta^2_t} (\beta_t \partial_z\psi + \beta_z\partial_t\psi)
\big] \Upsilon^2 \big\} \nonumber\\
&-& g(sA_t+cZ_t)\big\{ {\beta_t\Upsilon\Upsilon'\over
\sqrt{\beta^2_z+\beta^2_t}} + \big[ \partial_t\psi + {\beta_z \over
\beta^2_z+\beta^2_t} (\beta_t \partial_z\psi + \beta_z\partial_t\psi)
\big] \Upsilon^2 \big\}\nonumber\\
&-&{g^2\Upsilon^2\over 2(\beta^2_z+\beta^2_t)} \big[ (sA_z+cZ_z)\beta_z -
(sA_t+cZ_t) \beta_t \big]^2 + {g\Upsilon^2\over
\sqrt{\beta^2_z+\beta^2_t}} \big[ (sA'_z+cZ'_z)\beta_z +
(sA'_t+cZ'_t) \beta_t \big] \Big\} ,\\
\label{Ttilde}\end{eqnarray}

\noindent where the quotes [and the subsequent tildes in
Eqs.~(\ref{Utilde}) and~(\ref{Ttilde})] in the previous denominations
for the energy per unit length and tension come from the fact that the
stress tensor also has a nondiagonal part
\begin{eqnarray} A \equiv &\int& d^2x\,T^{zt} = 2\pi \int r\,dr\,\Big\{
-{g^2 \over 4c^2} h^2 Z_tZ_z - {g^2\over 4} h^2 (W^+_t W^-_z + W^+_z
W^-_t) - A'_t A'_z - Z'_tZ'_z - W^+_{rt} W^-_{rz} - W^+_{rz} W^-_{rt}
\nonumber \\
&-&ig\Big[(sA_z+cZ_z)(W^-_{rt} W^+_r - W^+ _{rt} W^- _r ) +
(sA_t+cZ_t)(W^-_{rz} W^+_r - W^+ _{rz} W^- _r ) \nonumber\\
&\quad &\qquad\qquad +(sA'_z+cZ'_z)(W^-_r W^+_t - W^+_r W^- _t ) +
(sA'_t+cZ'_t)(W^-_r W^+_z - W^+_r W^-_z ) \Big] \nonumber\\
&+&g^2\Big[ (W^+_r)^2 W^-_t W^-_z + (W^-_r)^2 W^+_t W^+_z - |W_r|^2
(W^+_z W^-_t + W^+_t W^-_z ) - 2 |W_r|^2 (sA_z+cZ_z)(sA_t+cZ_t)\Big]
\Big\}
,\end{eqnarray}

\noindent and this part can be removed by choosing the reference frame
in which either $\beta_t$ or $\beta_z$ vanishes, which is always
possible provided they are not equal, i.e., provided the current is
not lightlike. In this case however, the eigenvectors $u^\mu$ and
$v^\mu$ of the stress energy tensor~(\ref{Tmn}) are both null and the
definition of the eigenvalues as energy per unit length and tension
becomes unclear. Investigation of such lightlike currents in strings
is posponed for future work~\cite{jgpp}.

Therefore, in general, as was already discussed in
Ref.~\cite{low-mass}, the knowledge of the energy per unit length and
the tension in the particular cases $\beta _t = 0$, $\beta _z >0$ and
$\beta _z = 0$, $\beta _t >0$ will be sufficient for macroscopic
applications. For these two cases, it turns out that $U$ and $T$
(without the tilde, i.e., after diagonalization of the stress tensor,
this operation corresponding to choosing [$\beta_z = 0$, $\partial_z
\psi =0$] or [$\beta_t = 0$, $\partial_t\psi =0$]) actually depend on
the gauge fields $A_\mu$ and $Z_\mu$ only through the new field
functions
\begin{equation} \nu P(r) = \partial _a \psi + g(sA_a + c Z_a)
,\end{equation}

\noindent and the orthogonal field
\begin{equation} \nu' R(r) = -{c\over s} \partial _a \psi + g(sZ_a - c A_a)
,\end{equation}

\noindent where $a$ denotes either $z$ or $t$ depending on to whether
one considers a magnetic or an electric state respectively, and where
the parameter $\nu'$ is not in fact arbitrary since it is given by
$\nu'=-c\nu /s$. As a result, the realistic string model given by the
Lagrangian~(\ref{lag}) is nearly as simple as the
Witten~\cite{witten,num-witt} bosonic toy model in the sense that out
of the initial 18 field functions ($\varphi$, $h$, $C_\mu$,
$W^\pm_\mu$, $A_\mu$, and $R_\mu$), the internal microscopic structure
is fully determined by the knowledge of 6 field functions [$\varphi (r)$,
$h(r)$, $Q(r)$, $\Upsilon (r)$, $P(r)$ and $R(r)$] and two free
parameters, namely the winding number $n$ and the state parameter
$\nu$. This is to be compared to the original Witten bosonic model
whose structure needs the knowledge of already 4 field functions and
the same two free parameters, and with very similar field equations.
Thus we believe that, apart from the spontaneous current generation
mechanism discussed in the present article, most qualitative
conclusions regarding this simple model should apply as well to this
more realistic model.

\section*{Conclusions.}

In examining the internal structure of cosmic strings arising in the
most simple string-forming extension of the standard electroweak
model, we have found that, because of the nonabelian nature of
$SU(2)\times U(1)$, the field $W$ can condense spontaneously in a
strings core if the coupling constant between the string-forming Higgs
field and the usual $SU(2)$ doublet Higgs is less than a critical
value. This phenomenon can be understood in the following way: for
certain values of the coupling constants between the string forming
Higgs field $\Phi$ and the $SU(2)$ doublet Higgs field $H$, the latter
has a vanishing VEV close to the strings core, so the initial
$SU(2)\times U(1)$ symmetry is restored. Therefore, the
intermediate vector bosons, just like the photon, remains massless in
this region. Any exitation of the photon field will thus have enough
energy to generate a pair $W^+ W^-$ through vacuum fluctuations. In
turn, the effectively massless $W$ particle, being charged, is
responsible for the existence of a nonvanishing electromagnetic field.
This turns out to be in fact an unstable fluctuation mode, and nonzero
VEV for $W$ and the photon therefore build up spontaneously.

By using the field equations for the electroweak fields in the
symmetry restored region, we have been able to exhibit explicitely
this instability, and to estimate what we believe to be a lower bound
on the time necessary for the current to be generated. This timescale
is, as expected, independent of the underlying string parameters
provided the latter are such that the symmetry restoration mechanism
actually occurs. It should be remarked that because the phenomenon
here described is essentially electromagnetic and involves only the
$W$ particle, the timescale found could have been deduced on
dimensionnal ground, namely $\tau \sim (eM_W)^{-1}$. Although this is
a huge time compared to the characteristic length of the string if the
underlying string forming theory is at GUT scale, it is still
sufficiently short to be irrelevant for cosmological considerations.
Thus, current-carrying strings seem quite generic in string-forming
GUT models since the potential~(\ref{pot}) is in fact very general
even as a low energy limit and, as we have seen, the current formation
mechanism is independent of the background string structure.

Considering the results of Ref.~\cite{low-mass} and the present
calculation, it can be concluded that for any value value of the
coupling between the string forming theory and the electroweak fields,
the resulting strings are superconducting in the sense of Witten,
whether the current builds up through tunneling (high frequencies
metastability~\cite{low-mass}) or instability. Thus, if cosmic string
exist, and if they are not arbitrarily decoupled from the low energy
physics (a requirement of ``naturalness''), then, they are
superconducting. Since the present knowledge in high energy physics
tells us that approximately half of the plausible GUT theories
contains cosmic strings, it means that we can estimate the existence
probability of superconducting cosmic string to be also of the order
1/2.

A final remark seems appropriate at this point: the current generation
we have exhibited here relies in fact entirely on the nonabelian
nature of $SU(2)$. This means that for a string forming GUT, it will
exist as well since GUT models usually involve large unifying groups
with nonabelian couplings, and various Higgses. Therefore, gauge
bosons having masses of the order of the GUT scale should
spontaneously condense in the string core, within a timescale given
this time by $\tau \sim (g M_{GUT})^{-1}$, where $g$ is the GUT group
coupling constant. The cosmological relevance of such effects is then
obvious since many of these gauge bosons are responsible for baryon
number violation, so that in particular, these strings would enhance
the primordial baryon number asymmetry.

\section*{Acknowledgements}

I wish to thank B.~Carter, A.-C.~Davis, P.~Fayet, J.~Garriga, M.~Hindmarsh,
E.~P.~S.~Shellard and A.~Vilenkin for many interesting and improving
discussions. This work was supported by SERC grant \#~15091-AOZ-L9.

\vskip1cm
\begin{figure}
\caption{Field configuration around the vortex exhibiting the
SU(2)$\times$U(1) restoration for $\varphi < \varphi _{min}$. This
picture was drawn using a large value for $|f|$ in order to have a
large symmetry restoration region.}
\end{figure}

\end{document}